\documentclass[12pt]{article}
\usepackage[utf8]{inputenc}
\usepackage{lipsum}
\usepackage{hyperref}
\usepackage{enumitem}
\usepackage{float}
\usepackage{stmaryrd}
\usepackage{graphicx}
\usepackage{caption} 
\usepackage{calc}
\usepackage{mathrsfs}
\usepackage{fancyhdr}
\usepackage{tcolorbox}
\usepackage{hyperref}
\usepackage{wrapfig}
\usepackage{graphicx} 
\usepackage{amssymb}
\usepackage{amsmath}
\usepackage{multicol}
\usepackage{braket}
\usepackage{abstract}
\usepackage{multicol}
\usepackage{xcolor}
\usepackage{appendix}
\usepackage{cite}
\title{\normalsize\bf Dynamical Selection of Horizon-BMS Goldstone Modes\\ in  Evaporating Black Holes}
\author{\normalsize {\sc Nihar Ranjan Ghosh\footnote{\tt g.nihar@iitg.ac.in}\ \ and Malay K. Nandy\footnote{\tt mknandy@iitg.ac.in {\rm (Corresponding Author)}}}\\
\normalsize \em Department of Physics, Indian Institute of Technology Guwahati\\
\normalsize \em Guwahati 781 039, India}
\date{September 3, 2026}

\begin{document}

\maketitle

\begin{abstract}
We investigate the dynamics of horizon soft degrees of freedom associated with near-horizon BMS supertranslations in an evaporating Vaidya-Schwarzschild black hole spacetime. Considering the dynamic nature of the supertranslation parameter, we derive its effective action directly from the Einstein-Hilbert action. The resulting Goldstone sector is intrinsically coupled to the evolving black hole dynamics, with the mass function entering the evolution of the Goldstone modes while the Goldstone configuration contributes to the dynamical evolution of the black hole mass. Azimuthal periodicity and regularity at the poles select the physically admissible angular sector and yields a mass-dependent {\em selection rule} on the azimuthal modes. Consequently, the spectrum of horizon-supported Goldstone modes evolves with the black hole mass; as the horizon shrinks during evaporation, the horizon is left with a progressively reduced set of lower-order modes. The evaporation thus induces a dynamical filtering of the horizon soft sector, arising intrinsically from the black hole mass dynamics. These results provide an effective gravitational framework linking near-horizon BMS symmetry, dynamical Goldstone modes, and black hole evaporation,  suggesting a direct connection between macroscopic horizon evolution with the microscopic organization of the gravitational soft degrees of freedom residing on the horizon.
\end{abstract}

\maketitle
\tableofcontents
\section{Introduction}
Black holes provide a unique platform in which the fundamental aspects of gravity, thermodynamics, and quantum theory come together. The geometric description of a black hole is intrinsically tied to its thermodynamic properties, most notably through the discovery of Bekenstein and Hawking \cite{PhysRevD.7.2333, hawking1975particle}, relating the horizon area to the black hole entropy and the emission of Hawking radiation \cite{PhysRevD.14.2460}. These remarkable results indicate that the horizon is not merely a geometric boundary, but also signifies non-trivial degrees of freedom associated with the underlying gravitational dynamics. Understanding the nature of these degrees of freedom and their relation to the symmetries of spacetime is therefore an important direction in the study of black hole physics.

Hawking evaporation of the black hole seems to imply destruction of information when a black hole evaporates following its formation \cite{Hawking:1975vcx, hawking1974black}, indicating a non-unitary evolution from a pure state to a mixed state. This leads to what is called the information paradox, since such an evolution is forbidden in quantum mechanics. Even though several attempts have been made to resolve this paradox both in semi-classical and quantum gravity, the first possible resolution came through the realization that black holes can carry an infinite number of soft gravitational charges \cite{PhysRevLett.116.231301}. These infinite number of {\em constrained}  charges \cite{Ghosh:2026cno} are the symmetry charges corresponding to the Bondi--van der Burg--Metzner--Sachs (BMS) symmetry group \cite{bondi1962gravitational,sachs1962gravitational}, which extends the Poincar\'e group of asymptotically flat spacetimes by an infinite-dimensional set of supertranslations. Presence of these additional symmetry charges suggests that the gravitational phase space contains soft degrees of freedom which, although carrying arbitrarily low energy, can characterize non-trivial information about the state of the black hole.

The BMS symmetry, originally formulated at null infinity, has consequently provided a useful framework for investigating its relation with black hole thermodynamics \cite{Haco_2018,koga,Iofa}, soft theorems \cite{  Hawking:2016sgy, haco2018black, PhysRevD.96.084032, Chu_2018, PhysRevD.108.044034, strominger2017blackholeinformationrevisited, haco2019kerrnewmanblackholeentropy, PhysRevD.103.126020, Donnelly:2014fua}, gravitational memory effect ~\cite{Strominger:2014pwa, strominger2018lecturesinfraredstructuregravity, PhysRevD.92.084057, Flanagan:2015pxa, Pasterski:2015tva, Pasterski:2015zua, PhysRevD.98.064032, Compere:2016jwb, Bieri:2013hqa, susskind2019electromagneticmemory, PhysRevLett.116.091101, PhysRevD.102.044041}, and Hawking radiation \cite{chu2018soft,nihar:entropy}. Supertranslations correspond to angle-dependent translations and generate distinct configurations in the gravitational phase space\cite{strominger2018lecturesinfraredstructuregravity}. In the context of black holes, this structure suggests that the horizon itself can support non-trivial transformations at the horizon \cite{JabbariPRL,JabbariJHEP} analogous to BMS transformations in the  asymptotic limit, $r\to\infty$. When such {\em large} diffeomorphisms act non-trivially at the horizon, they cannot be regarded merely as gauge redundancies; instead, their action gives rise to physical boundary degrees of freedom \cite{PhysRevLett.116.231301,haco2018black}. This provides a natural connection between the soft sector associated with BMS symmetry and the degrees of freedom localized on the black hole horizon. In particular, the horizon can be viewed as a boundary supporting soft gravitational excitations whose dynamics is expected to become relevant when the black hole evolves.

An important realization of this picture is obtained by interpreting the horizon supertranslation modes as Goldstone modes associated with spontaneous breaking of the BMS symmetry. A black hole background selects a particular configuration from the family of geometries related by such supertranslations. The transformation parameter connecting these configurations can then acquire a physical meaning and be promoted to a dynamical field on the horizon \cite{Averin_2016}. Thus, the Goldstone field is not introduced as an independent matter degree of freedom; rather, it emerges from the non-trivial action of the horizon supertranslation symmetry of the gravitational configurations. This viewpoint is closely related to the interpretation of soft modes as edge degrees of freedom associated with large gauge transformations acting at the boundary
\cite{Donnelly:2014fua,Donnelly:2015hxa,Harlow:2015lma,Harlow:2016vwg}. For black holes, the horizon therefore provides a natural framework in which the spontaneous breaking of the BMS supertranslation symmetry can generate physical Goldstone excitations \cite{Eling:2016xlx,averin2016gravitational,mpla,PhysRevLett.116.231301,hawking2017superrotation,MAITRA2022136825}.

The relevance of these horizon Goldstone modes becomes particularly interesting for dynamical black holes. For a stationary black hole, the horizon geometry is largely characterized by a small number of macroscopic parameters (such as, mass, electric charge and angular momentum), and the action of a supertranslation generates angle dependent horizon degrees of freedom that remain stationary. Once the horizon is allowed to evolve, however, the situation becomes considerably richer, since these horizon soft modes also evolve  with the evolving mass function. These modes can provide a dynamical realization of the soft sector. Consequently, studying the Goldstone modes together with the black hole evolution can offer a way of relating the horizon symmetry structure to both the thermodynamic properties of the black hole and the radiation produced during its evaporation.

In this paper, we consider a Vaidya-Schwarzschild black hole and investigate the dynamics of the Goldstone modes associated with horizon supertranslations. Treating the event horizon as a boundary on which the BMS transformations act non-trivially \cite{koga, Iofa}, we construct the corresponding effective action and study the coupled evolution of these modes with the background geometry. In a recent study \cite{nihar:entropy}, by considering the surface sector of the horizon-BMS Goldstone action, a direct connection between these modes and the Bekenstein-Hawking entropy was established. While the leading-order contribution reproduces the Bekenstein-Hawking area law, $S=A/4$, the Goldstone modes generate non-trivial sub-leading corrections whose sign depends on their dynamics. This provides a preliminary indication that the black hole entropy carries a dependency on the horizon-BMS sector. To quantify this dependence and understand its physical origin, it is therefore necessary to determine the dynamics of the Goldstone fields explicitly. Since the surface sector primarily governs the entropy contribution, we focus here on the bulk sector, which provides the dynamical equations for the Goldstone modes. We construct the bulk action within an effective field theory framework and subsequently obtain analytical solutions for the coupled dynamics. The resulting analysis provides a dynamical description of the horizon-BMS Goldstone sector and its interaction with the evolving black hole geometry.

The physical picture underlying our construction is that the Vaidya-Schwarzschild geometry is acted upon by a BMS-generating vector field, whose non-trivial action modifies the near-horizon geometry and breaks its spherical symmetry. The associated spontaneous breaking of horizon supertranslation invariance gives rise to Goldstone modes localized on the horizon, with dependence on both the advanced time and the angular coordinates. These modes are identified with the supertranslation parameter and therefore characterise the fluctuations associated with the broken horizon symmetry. The Vaidya background, on the other hand, provides a natural setting in which the black hole mass evolves dynamically and describes the evaporation process. The resulting system thus allows the temporal evolution of the horizon Goldstone modes together with the changing black hole mass. Exploring this coupled dynamics can consequently provide further insight into how horizon-BMS symmetry and its associated Goldstone degrees of freedom respond to, and potentially influence, the evolution of a dynamical black hole.

The rest of the paper is organized as follows. In Sec.~\ref{sec near horizon}, we introduce the near-horizon BMS symmetry of the Vaidya-Schwarzschild black hole and discuss the emergence of the associated Goldstone modes. In Sec.~\ref{sec action}, we construct the effective Goldstone action from the Einstein-Hilbert action and derive the corresponding equations of motion. Sec.~\ref{sec sol} is devoted to solving the Goldstone field equations and determining their coupled dynamics with the black hole mass function. We first analyze the regularity conditions associated with two possible ans\"atze for the Goldstone field and subsequently obtain the analytical solution for the evolving black hole mass. Finally, in Sec.~\ref{disc}, we discuss the physical implications of our results and their relevance to the interplay between horizon-BMS symmetry, Goldstone modes, and black hole dynamics.

\section{Near Horizon BMS Symmetry}\label{sec near horizon}
Since we are interested in the near horizon analysis, where the Goldstone fields reside, it will be convenient to express the Schwarzschild metric in terms of a coordinate system which is particularly well suited in the near horizon limit. One of such possible, well suited coordinate system is the Rindler coordinate system.  The metric of a static Schwarzschild black hole, expressed in a near horizon Rindler coordinate system, is given by  
\begin{equation}
    \label{rindler coordinate}
    ds^2=-\frac{\rho^2}{16m^2}dv^2+\frac{\rho}{2m}dvd\rho+4m^2d\Omega_2^2~,
\end{equation}
where $m$ is  the constant mass of the black hole, $d\Omega_2^2=d\theta^2+\sin^2\theta d\phi^2$ is the metric of a $2$-sphere. The horizon of the black hole is located at $\rho=0$. 

Defining a new radial coordinate as $r=\rho^2/(8m)$, we have from equation \ref{rindler coordinate},
\begin{equation}
    \label{metric}
    ds^2\Bigg|_{g_{ab}}=-\frac{r}{2m}dv^2+2dvdr+4m^2d\Omega_2^2~.
\end{equation}
Although the near-horizon geometry in stationary black hole settings has been explored extensively, its application for dynamical horizons remain much less studied. Moreover, to associate the supertranslation parameter (Goldstone modes) with the black hole evaporation, a dynamical spacetime is needed. The Vaidya–Schwarzschild geometry provides one of the simplest and most useful models of the non-stationary black hole, describing evaporation through a time-dependent mass function. This geometry is obtained with the transformation $m\to m(v)$. Thus, the near horizon metric for a Vaidya Schwarzschild black hole, can be written as
\begin{equation}
    \label{original metric}
    ds^2\Bigg|_{g_{ab}}=-\frac{r}{2m(v)}dv^2+2dvdr+4m(v)^2d\Omega_2^2~.
\end{equation}
Interestingly, the volume element of the $v$-$r$ constant hypersurface is now time dependent, which is expected as the metric represents the near horizon $(r=0)$ geometry of a Schwarzschild-Vaidya black hole. As the black hole evaporates or accumulates mass, its horizon radius changes and so does the volume element on the horizon.

As mentioned earlier, we choose a particular set of gauge conditions which preserves the near horizon structure of the black hole in \ref{original metric}. One such set of appropriate gauge choices is as follows: 
\begin{equation}
    \label{Lie-eta}
    \mathcal{L}_{\eta}g_{rr}=\mathcal{L}_{\eta}g_{rv}=\mathcal{L}_{\eta}g_{Ar}=0~~.
\end{equation}
The vector field $\eta$, which satisfies the above gauge conditions in equation \ref{Lie-eta}, is given by 
\begin{equation}
    \label{eta}
    \eta=F\partial_v-r\partial_vF\partial_r-\frac{r}{4m^2 \gamma_{AA}}\partial_A~~,
\end{equation}
where $A=\theta,\phi$, with $\gamma_{AA}$ the $AA^{th}$ element of the $2$-sphere metric. The arbitrary function $F(v,\theta,\phi)$ represents the supertranslation parameter. 

The Goldstone theorem states that whenever a continuous global symmetry is spontaneously broken, there occurs a massless excitation about the spontaneously broken vacuum, known as Goldstone boson. In the context of BMS group, the horizon supertranslation symmetry is spontaneously broken due to the action of the vector field $\eta$, so that $F(v,\theta,\phi)$ appears as a Goldstone mode living on the horizon. Although $F$ originates as a large gauge parameter, it parametrizes inequivalent near-horizon geometries (differing by soft hair) and can carry physical information \cite{Lin_2022}.

Moreover, unlike the asymptotic $r\to\infty$ BMS analysis, where the supertranslation parameter is a function of the angular parameters only, here in our near-horizon analysis, the supertranslation parameter $F(v,\theta,\phi) $ is also a function of the time coordinate $v$ which is a direct consequence of the chosen gauge conditions. This property further enables us to treat $F(v,\theta,\phi) $ as the Goldstone boson of the spontaneously broken symmetry and to explore its evolution in time, rather than confining on a constant $v$-$r$ hypersurface. This is essential if one hopes to connect horizon symmetries to dynamical black holes with $m=m(v)$. Time dependent supertranslation parameter, in the context of horizon BMS symmetry, has also been obtained in other analyses as well \cite{koga,MAITRA2022136825}.

As will be shown in the next part, under the action of the vector field \ref{eta} that generates the {\em large diffeomorphism}, the underlying gravitational field $g_{ab}$ will be modified. This can be thought of as similar to the transformation which breaks the U$(1)$ symmetry of a scalar field $\Phi$. The modification in the geometry is parametrized by the supertranslation parameter $F(v,\theta,\phi)$. Consequently, the macroscopic parameters of the original metric will be modified, and therefore with the analogy of U$(1)$ symmetry breaking, this can be regarded as breaking of the horizon supertranslation symmetry. Therefore, we can interpret the parameter $F$ as Goldstone mode. Moreover, in analogy with the U(1) Goldstone mode action which is obtained from the scalar field action, we shall obtain in our case the action for the Goldstone field $F$ from the Einstein–Hilbert action.

As a result of the supertranslation of $g_{ab}$ generated by the vector field $\eta$, given by \ref{eta}, the modified metric is obtained as
\begin{equation}
    \label{new metric}
    \begin{split}
        ds^2\Bigg|_{\bar{g}_{ab}}&=g_{ab}+\mathcal{L}_{\eta}g_{ab}\\
        &=r\left(-\frac{1}{2m}+\mathcal{F}_1\right)dv^2+2dvdr+r\partial_A\mathcal{F}_2dvdx^A+\left(4m^2+8mm'F-2r\partial_\theta^2F\right)d\theta^2\\
        &-2r(\partial_\theta\partial_\phi F-\cot\theta\partial_\phi F)d\theta d\phi+\left[(4m^2+8mm'F)\sin^2\theta-2r(\sin\theta\cos\theta\partial_\theta F+\partial_\phi^2F)  \right]d\phi^2~,
    \end{split}
\end{equation}
where
\begin{equation}
    \label{F1}
    \mathcal{F}_1=\frac{m'}{2m^2}-\frac{1}{2m}\partial_v F-2\partial_v^2F~,
\end{equation}
and
\begin{equation}
    \label{F2}
    \mathcal{F}_2=\frac{2m'}{m}F-\frac{1}{2m}F-2\partial_vF~.
\end{equation}
 
It is important to note that, because the original unperturbed metric is of the Vaidya type, the near horizon BMS transformation acts differently compared to the static background case. The metric perturbation $h_{ab}$, generated by the action of the diffeomorphism generating vector field $\eta$, and parametrized by the supertranslation $F(v,\theta,\phi)$, turns out to be
\begin{equation}
    \label{hab}
h_{ab}=\mathcal{L}_{\eta}g_{ab}=\begin{bmatrix}
r\mathcal{F}_1 & 0 & \frac{r}{2}\partial_\theta\mathcal{F}_2 & \frac{r}{2}\partial_\phi\mathcal{F}_2  \\
0 & 0 & 0 & 0  \\
\frac{r}{2}\partial_\theta\mathcal{F}_2 & 0 & \bar{g}_{\theta\theta}-4m^2 &  \bar{g}_{\theta\phi} \\
\frac{r}{2}\partial_\phi\mathcal{F}_2 & 0 & \bar{g}_{\theta\phi} & \bar{g}_{\phi\phi}-4m^2\sin^2\theta\\
\end{bmatrix}~~,
\end{equation}
where $\bar{g}_{AB}$ can be found from equation \ref{new metric}. 

Notably, the action of the vector field $\eta$ breaks the global SO$(3)$ symmetry, as is clear from equation \ref{new metric}. This particular $h_{ab}$, parametrized by the supertranslation parameter $F(v,\theta,\phi)$, will be used in the next section to obtain the effective action and the energy momentum tensor of the Goldstone field.

\section{Effective Goldstone Action and Equations of Motion}\label{sec action}
In the preceding section, $\bar{g}_{ab}$ was introduced as the transformed Vaidya metric obtained through the action of the near-horizon BMS symmetry. As evident from its explicit form in Eq.~\ref{new metric}, the Goldstone field $F$ enters the metric in a highly non-linear manner, rendering a direct analysis of the corresponding dynamics mathematically cumbersome. Furthermore, as discussed above, we aim to derive the effective action for the Goldstone field solely from the Einstein-Hilbert action, following the standard approach of constructing effective theories from the underlying fundamental action. This will be achieved by treating the metric perturbation induced by the BMS transformation, $h_{ab}=\mathcal{L}_{\eta}g_{ab}$, as a small perturbation around the original background metric $g_{ab}$. The Einstein-Hilbert action can then be systematically expanded in powers of $h_{ab}$, or equivalently in the Goldstone field, and truncated at the desired order. This perturbative expansion provides a tractable framework for extracting the effective dynamics of the horizon Goldstone modes directly from the Einstein-Hilbert action.

The Einstein-Hilbert action, written in terms of the  new Ricci scalar $\bar{R} $, derived from the metric $\bar{g}_{ab} $, is expressed as 
\begin{equation}
    \label{EH action}
    S_E=\int d^4x\sqrt{-\bar{g}}~\bar{R}~,
\end{equation}
with $16\pi G=1$.

Considering  $h_{ab}=\mathcal{L}_{\eta}g_{ab}$ as small fluctuations on the background metric $g_{ab}$, the Taylor series expansion of equation \ref{EH action} is given by 
\begin{equation}
    \label{taylor}
    S_E\Bigg|_{\bar{g}_{ab}}=S_E[g_{ab}]+h_{ab}\left(\frac{\delta S_E}{\delta \bar{g}_{ab}} \right)\Bigg|_{g_{ab}}+h_{ab}h_{cd}\left(\frac{\delta^2 S_E}{\delta \bar{g}_{ab}\delta \bar{g}_{cd}}  \right)\Bigg|_{g_{ab}}+\dots
\end{equation}
The first term on the right hand side does not contribute to the analysis. Moreover, the second term is proportional to the Einstein tensor that vanishes  when the background metric $g_{ab}$ is a vacuum solution of the Einstein's equations of motion. (In any non-trivial case with a non-zero energy momentum tensor $T_{ab }$, the second term gives the field equations for the metric field $g_{ab} $.) Importantly, the field equation for the Goldstone field $F$ originates from the third term that governs fluctuations around the vacuum at the horizon. This is analogous to Goldstone excitations in spontaneous symmetry breaking as pointed out earlier. Analogously, $F$ describes low-energy excitations around the vacuum solution at the horizon. Thus, the third term is the action of the corresponding Goldstone mode $F$. Higher power terms in $h_{ab}$ are expected to contribute in the sub-leading order as we treat $h_{ab}=\mathcal{L}_{\eta}g_{ab}$ as small fluctuations on the background spacetime $g_{ab}$.

Considering the third term $\mathcal{O}(h^2)$ in equation \ref{taylor}, and using the toolkit developed in \cite{Ghosh:2026cno}, the bulk action for the Goldstone field turns out to be
\begin{equation}
    \label{goldstone action}
    \begin{split}
        S_F=\frac{1}{2}M_p^2\int& d^4x\sqrt{-g}\Big[ h^{ab} \square h_ {ab}+h^{ab} \nabla_a \nabla_b h+\frac{1}{2}h \nabla_a \nabla_b h^{ab}-\frac{1}{4}h\square h+\frac{3}{4}\nabla_a h_{bc}\nabla^a h^{bc}\\
        &+\frac{1}{2}\nabla_b h_{ac}\nabla^a h^{bc}+R^{ab}h_{ac}h^c_b+\frac{R}{8}(h^2-2h^2_{ab})-\frac{1}{2}h_{ab}R^{ab}\Big],
    \end{split}
\end{equation}
where $h^2=(g^{ab}h_{ab})^2$ and $h^2_{ab}=h^{ab}h_{ab} $, and $R$, $R_{ab}$ are with respect to the background metric $g_{ab}$, and the $\nabla$'s are with respect to the unperturbed metric $g_{ab} $ in equation \ref{original metric}. 

Having obtained the scalar field action $S_F$ for the Goldstone field $F$, the corresponding energy momentum tensor $T_{\mu\nu}$ is obtained by varying the action $S_F$ with respect to the metric field $g_{ab}$ by treating $ h_{ab}$ as an independent field. This yields the energy momentum tensor for the Goldstone field as
\begin{equation}
    \label{energy momentum tensor}
    \begin{split}
       8 T_{\mu\nu}&=g_{\mu\nu}\Big[Rh^2-4h R_{ab}h^{ab}-2\nabla_ah\nabla^ah-16\nabla_ch_b^c\nabla^{[a}h_a^{b]}-4\nabla_bh_{ac}\nabla^ch^{ab}+6\nabla_ch_{ab}\nabla^ch^{ab}\\    &+8h\nabla_c\nabla^{[b}\nabla_b^{c]}+2h_{ab}\Big\{4h^c_aR_{bc}-Rh_{ab}+8\nabla_b\nabla_ah-4\nabla_b\nabla_ch_a^c-4\nabla_c\nabla_bh_a^c+4\square h_{ab}    \Big\} \Big]\\
       &-24\nabla_b\nabla_ah^{ab}h_{\mu\nu}+16\nabla^bh_\mu^a\nabla_{[a}h_{b]\nu}+8\nabla_ah^a_\nu\nabla_\mu h-4\nabla_\mu h^{ab}\nabla_\nu h_{ab}+8\nabla_ah^a_\mu \nabla_\nu h\\
       &+8\nabla_bh^b_a\Big(\nabla_\mu h^a_\nu+\nabla_{\nu}h^a_{\mu}-\nabla^a h_{\mu\nu} \Big)-12\nabla_ah\nabla^{(a}h_{\mu\nu)}-4h^a_\mu\Big(4h^b_\nu R_{ab}-2h_{a\nu}R\Big)\\
       &4h^{ab}\Big[2h_{\mu\nu}R_{ab}+h_{ab}R_{\mu\nu}+2\nabla_b\Big(\nabla_\mu h_{a\nu}+\nabla_{\nu}h_{a\mu}-\nabla_a h_{\mu\nu} \Big)-2\nabla_{(\mu}\nabla_{\nu)}h_{ab}   \Big]-2h\Big[hR_{\mu\nu}\\
       &+2h_{\mu\nu}R-\square h_{\mu\nu}+4\nabla_b\nabla_{(\mu}h_{\nu)^b}-2\nabla_{(\mu}\nabla_{\nu)}h\Big]+\Big[\tau_{\mu\nu}+(\mu\leftrightarrow\nu) \Big]~,
    \end{split}
\end{equation}
with $2A_{(a}B_{b)}=A_aB_b+A_bB_a$, $2A_{[a}B_{b]}=A_aB_b-A_bB_a$ and
\begin{equation}
    \label{tau}
    \begin{split}
        \tau_{\mu\nu}=&4h_\mu^a\Big[2hR_{\nu a}-4h^b_aR_{\nu b}-\nabla_a\nabla_\nu h+2\nabla_b\Big(\nabla_a h^b_\nu+\nabla_{\nu}h^b_{a}-\nabla^b h_{a\nu}\Big)-3\nabla_\nu\nabla_ah\Big]~.
    \end{split}
\end{equation} 

Using the expression for $h_{ab}$ from equation \ref{hab} in the Goldstone action \ref{goldstone action}, the equation of motion for the Goldstone field can be obtained from the Euler-Lagrange equation for higher order derivative theories \cite{goldstein2011classical}
 
\begin{equation}
    \label{Euler-Lagrange}
    \left[\frac{\partial \mathcal{L}_F}{\partial F}-\partial_a\left(\frac{\partial \mathcal{L}_F}{\partial(\partial_a F)}  \right)+\partial_a\partial_b\left(\frac{\partial \mathcal{L}_F}{\partial(\partial_a\partial_bF)}  \right)\right]\Bigg|_{r=0}=0~.
\end{equation}
Thus, the equation of motion for the Goldstone field on the horizon $r=0$  turns out to be
\begin{equation}
    \label{KG eq}
    \begin{split}
        &16 m \left[m'' \partial_\theta^2F+m\partial_v^2\partial_\theta^2 F\right]+64 m'^2\partial_\theta^2 F-4 m' \left[8 m\partial_v\partial_\theta^2 F+17\partial_\theta^2 F\right]\\
        &+32 \cot \theta  \Big[\left\{3 m'^2-m m''\right\}\partial_\theta F-2 m m'\partial_v\partial_\theta F\Big]\\
        &+\csc ^2\theta \Big[16 m \left\{2 m' \left(\partial_vF-\partial_v\partial_\phi^2F\right)+m\partial_v^2\partial_\phi^2 F\right\}+\Big(16 m m''+64 m'^2\\
        &-68 m'-1\Big)\partial_{\phi}^2 F+32 \left\{m m''+m' \left(3 m'-1\right)\right\} F\Big]-\partial_{\phi}^2F=0~.
    \end{split}
\end{equation}

The ``00" component of the Einstein tensor, following from the metric \ref{original metric}, on the horizon $r=0$ is given by
\begin{equation}
    \label{einstein}
    \begin{split}
        G_{00}&=\frac{-4mm''+m'}{2m^2}~.
    \end{split}
\end{equation}
Consequently, if we consider the unperturbed background metric $g_{ab} $ to be static, the second term in the Taylor series expansion in equation \ref{taylor}, which is proportional to the Einstein tensor, gives zero contribution and one is only left with the Goldstone field equation of motion.

Similar to the derivation of Goldstone field equation of motion, by evaluating the $\nabla$'s with respect to the unperturbed metric $g_{ab}$, the ``00" component of the Einstein field equation $G_{00}=T_{00}$ on the horizon $r=0$, turns out to be
\begin{equation}
    \label{00 enst eqn}
    \begin{split}
        &(4mm''-m')(2m^2)=4 m m' \partial_v F \left[m' \left(4 m\partial_v F-3\right)+3 m \left(4 m \partial_v^2F+\partial_vF\right)\right]\\
        &+F \Big[12 m^2 m'' \left(4 m\partial_v^2 F+\partial_vF\right)+m'^2 \left\{4 m \left(5 \partial_vF-8 m\partial_v^2 F\right)-3\right\}\\
        &+m m' \left\{\left(3-32 m m''\right)\partial_v F-48 m^2 \partial_v^3F\right\}+8 m'^3 \left(4 m\partial_v F-3\right)\Big]\\
        &+8 \Big[2 m^2 m''^2+6 m'^4+4 m m'^2 m''+m m' \left(m''-4 m \partial_v{^3} m\right)\Big] F^2
    \end{split}
\end{equation}

Thus, we finally have two coupled nonlinear partial differential equations, given by \ref{KG eq} and \ref{00 enst eqn}, for the black hole mass function $m(v)$ and the Goldstone field $F(v,\theta,\phi)$. A notable feature of the resulting equations is that the horizon Goldstone sector and the black hole evolution are dynamically connected intimately. Equation \eqref{KG eq} determines the evolution of the Goldstone field in response to the time-dependent background, while equation \eqref{00 enst eqn} contains explicit contributions from the Goldstone configuration and consequently determines the evolution of the black-hole mass. The horizon supertranslation modes therefore act as a genuine dynamical sector coupled to the evaporation dynamics, rather than as a purely kinematical parametrization of the horizon modes. This provides an effective description in which the evolution of the black hole background and that of its horizon soft degrees of freedom are determined self-consistently.

\section{Solution for Goldstone Field and Black Hole Mass}\label{sec sol}
The nonlinear partial differential equations \ref{KG eq} and \ref{00 enst eqn} are intricately coupled, and their exact solution is not readily apparent. Nevertheless, an approximate analytical treatment can provide valuable insight into the qualitative behaviour and the underlying behaviour of the solutions. To this end, we adopt a separable ans\"atz for the Goldstone field of the form $F(v,\theta,\phi)=f(v)\,g(\theta,\phi)$. Substituting this ansatz into Eq.~\ref{KG eq} yields
\begin{equation}
    \label{KG1}
    \begin{split}
        g f_1(v)+f_2(v)\left(\partial_\phi^2g+\sin^2\theta\partial_\theta^2g\right)-f_3(v)\sin\theta\cos\theta\partial_\theta g=0~,
    \end{split}
\end{equation}
where
\begin{equation}
    \label{KG coefficients}
    \begin{split}
        f_1(v)&=32\Big[m'\{mf'+f(3m'-1)\}+fmm''\Big],\\
        f_2(v)&=16m(mf''-2f'm')+f\{16mm''+64(m')^2-68m'-1   \},\\
        f_3(v)&=32\Big[ -3f(m')^2+m(2f'm'+fm'') \Big].
    \end{split}
\end{equation}
\subsection{Analytical solution for angular part of Goldstone field}
With further substitution of $g(\theta,\phi)=\Theta(\theta)\Phi(\phi)$, equation \ref{KG1} leads to
\begin{equation}
    \label{KG2}
    \frac{f_1}{f_2}+\frac{\Phi''}{\Phi}+\sin^2\theta\frac{\Theta''}{\Theta}-\frac{f_3}{f_2}\sin\theta\cos\theta\frac{\Theta'}{\Theta}=0~.
\end{equation}
Equation \ref{KG2} leads to the periodic solution for the azimuthal part as 
\begin{equation}
    \label{Phi}
    \Phi(\phi)=c_1e^{ik\phi}+c_2e^{-ik\phi}~,
\end{equation}
where the separation constant is constrained as $k=0,\pm1,\pm2\dots$, due to the periodicity condition $\Phi(\phi+2\pi)=\Phi(\phi)$.

Hence equation \ref{KG2} simplifies to
\begin{equation}
    \label{KG3}
    \frac{f_1}{f_2}+\sin^2\theta\frac{\Theta''}{\Theta}-\frac{f_3}{f_2}\sin\theta\cos\theta\frac{\Theta'}{\Theta}=k^2
\end{equation}

An exact analytical solution of this equation is generally intractable. Although numerical methods can provide solutions, they may obscure the underlying physical structure in terms of the functional trend of the solutions. We therefore seek an approximate analytical solution that captures the {\em essential physics}  of the coupled dynamics. In particular, we require the mass function $m(v)$ to decrease monotonically with the retarded time $v$, with $m(v)\to 0$ as $v\to\infty$, consistent with the expected late-time decay of the Goldstone modes. In addition, the angular function $\Theta(\theta)$ is required to remain regular and finite at the poles, $\theta=0,\pi$. These physical and regularity conditions provide appropriate constraints on the admissible solutions and are expected to capture the essential behaviour of the Goldstone modes and their backreaction on the black hole mass through the coupled dynamics.

Accordingly, to obtain an approximate analytical solution of Eq. \ref{KG3}, we consider the following two ans\"atze:

 \begin{enumerate}
    \item Ans\"atz I: 
    \begin{equation}
        \label{choice 1}
        f_1=0,~~ \text{and}~~f_3/f_2=\text{constant}=N
    \end{equation}
    \item Ans\"atz II:
    \begin{equation}
        \label{choice 2}
        f_3=0,~~ \text{and}~~f_1/f_2=\text{constant}=N~.
    \end{equation}
\end{enumerate} 

Next, we assess both ans\"atze for physical admissibility. If both ans\"atze yield physically consistent solutions, then each may encode relevant aspects of the underlying dynamics, and there would be no clear physical criterion for preferring one over the other. In such a situation, relying solely on the approximate ans\"atze would not be sufficient to identify the physical solution, and an exact or more systematic treatment of Eq. \ref{KG3} would be required. Conversely, if only one of the two ans\"atze produces a physically admissible solution, it is natural to regard that ans\"atz as capturing the essential physical behavior encoded in Eq. \ref{KG3}, while the other should be discarded on physical grounds.

The ans\"atz \ref{choice 1}, when substituted in equation \ref{KG3}, leads to the solution
\begin{equation}
    \label{theta}
   \begin{split}
       \Theta(\theta)&= c_3 (\sin\theta)^{(1 + N)/2}
   \text{LegendreP}[\frac{1}{2} (N-1), \frac{1}{2} \sqrt{1 + 2 N + N^2 + 4 k^2}, 
   \cos\theta]\\
   &+ 
 c_4 (\sin\theta)^{(1 + N)/2}
   \text{LegendreQ}[\frac{1}{2} (N-1),\frac{1}{2} \sqrt{1 + 2 N + N^2 + 4 k^2}, 
   \cos\theta]~,
   \end{split}
\end{equation}
where $\text{LegendreP}[\mu,\nu,x]$ and $\text{LegendreQ}[\mu,\nu,x]$ are the Legendre functions of the first and second kind, repectively.

On the other hand, substituting the alternative ans\"atz \ref{choice 2} in equation \ref{KG3} leads to
\begin{equation}
    \label{theta2}
   \begin{split}
       \Theta(\theta)&= c_5 (\sin\theta)^{1/2}
   \text{LegendreP}[\frac{1}{2}, \frac{1}{2} \sqrt{1 -4 N + 4 k^2}, 
   \cos\theta]\\
   &+ 
 c_6 (\sin\theta)^{1/2}
   \text{LegendreQ}[\frac{1}{2},\frac{1}{2} \sqrt{1 -4 N + 4 k^2}, 
   \cos\theta]~.
   \end{split}
\end{equation}

In the next subsection, we shall analyse the regularity behaviours of the above two solutions for $\Theta(\theta)$ at the poles $\theta=0,\pi $ in order to assess their physical admissibility.

\subsubsection{Regularity Conditions for Ansatz I }
To analyse the regularity of $\Theta(\theta)$ at $\theta=0,\pi$ for the solution in equation \ref{theta} obtained with ans\"atz \ref{choice 1}, we first choose $c_4=0$, leading to 
\begin{equation}
    \label{theta final}
   \begin{split}
       \Theta(x)&= c_3 (1-x^2)^{(1 + N)/4}
   \text{LegendreP}[\frac{1}{2} (N-1), \frac{1}{2} \sqrt{1 + 2 N + N^2 + 4 k^2}, 
   x],\\
   \end{split}
\end{equation}
where $x=\cos\theta$. Since $\text{LegendreP}[\nu,\mu,x] \equiv P_\nu^{\mu}(x)$, we first look at its behaviour near the poles, $x=\pm1$, corresponding to $\theta=0,\pi$.

For non-integer $\mu>0$, the asymptotic expressions for the Legendre function $P_\nu^\mu(x)$ are \cite{abramowitzmath}
\begin{align}
P_{\nu}^{\mu}(x)
&\sim
\frac{2^{\mu/2}}{\Gamma(1-\mu)}
(1-x)^{-\mu/2},
\qquad x\to 1^{-}\label{P non-int mu, x=+1},
\\[6pt]
P_{\nu}^{\mu}(x)
&\sim
\frac{2^{\mu/2}\Gamma(\mu)}
{\Gamma(1+\nu)\Gamma(-\nu)}
(1+x)^{-\mu/2} \label{P non-int mu, x=-1},
\qquad x\to -1^{+}.
\end{align}
In our case, from equation \ref{theta final}, $\mu=\frac{1}{2} \sqrt{1 + 2 N + N^2 + 4 k^2}>0$ and $\nu=\frac{1}{2} (N-1)$. Therefore,  from equations \ref{P non-int mu, x=+1} and \ref{P non-int mu, x=-1}, in the limit $x\to\pm 1$, equation \ref{theta final} leads to
\begin{align}
\Theta(x)
&\sim
\big(1+x\big)^{\frac{1}{4}(N+1)} \big(1-x\big)^{\frac{1}{4}\big[(N+1)-\sqrt{(N+1)^2+4k^2}  \big]}
,
\qquad x\to 1^{-}\label{non-int mu, x=+1},
\\[6pt]
\Theta(x)
&\sim   
\big(1-x\big)^{\frac{1}{4}(N+1)}\big(1+x\big)^{\frac{1}{4}\big[(N+1)-\sqrt{(N+1)^2+4k^2}  \big]} \label{non-int mu, x=-1},
\qquad x\to -1^{+}.
\end{align}

Regularity of both \ref{non-int mu, x=+1} and \ref{non-int mu, x=-1} requires $(N+1)>\sqrt{(N+1)^2+4k^2}$, leading to 
\begin{equation}
    \label{imaginary k}
    k^2<0~.
\end{equation} 
This implies $k=\text{imaginary}$. However, the periodicity condition of the azimuthal part $\Phi(\phi)$ in equation \ref{Phi} restricts $k\in \mathbb{Z}$. Thus, we must discard the possibility of $\mu$ to be a non-integer. 

Thus Ans\"atz I given by \ref{choice 1} is not an acceptable ans\"atz, since $\Theta(\theta)$  diverges at $\theta=0,\pi$ for $\mu=\text{non-integer}$.

Consequently, the only remaining possibility is that $\mu\in\mathbb{Z}^+$,  in which case the asymptotic expressions for the Legendre functions are \cite{abramowitzmath}
\begin{align}
P_{\nu}^{\mu}(x)
&\sim
(-1)^\mu
\frac{\Gamma(\nu+\mu+1)}
{\mu!\,\Gamma(\nu-\mu+1)}
\left(\frac{1-x}{2}\right)^{\mu/2}\label{P int mu, x=1},
&& x\to 1^{-},
\\[6pt]
P_{\nu}^{\mu}(x)
&\sim
\frac{\sin(\pi\nu)}{\pi}
\,\Gamma(\mu)
\left(\frac{2}{1+x}\right)^{\mu/2}\label{P int mu, x=-1},
&& x\to -1^{+},
\end{align}

Consequently, the asymptotic forms of $\Theta(x) $ near the poles of equation \ref{theta final} are
\begin{equation}
    \label{theta divergent x=1}
    \Theta(x)\sim \big(1+x\big)^{\frac{N+1}{4}}\big(1-x\big)^{\frac{1}{4}[N+1+\sqrt{(N+1)^2+4k^2}]},~~~x\to1^-,
\end{equation}
and 
\begin{equation}
    \label{theta divergent x=-1}
    \Theta(x)\sim \big(1-x\big)^{\frac{N+1}{4}}\big(1+x\big)^{\frac{1}{4}[N+1-\sqrt{(N+1)^2+4k^2}]},~~~x\to-1^+,
\end{equation}
Equation \ref{theta divergent x=1} is regular at the pole $x=1$. However, for the regularity of equation \ref{theta divergent x=-1} at $x=-1^+$, we must impose the condition $N+1-\sqrt{(N+1)^2+4k^2}>0$, which implies 
\begin{equation}
    \label{imaginary k 2}
    k^2<0~,
\end{equation}
demanding again $k$ to be imaginary, a requirement that goes against the periodicity condition of $\Phi(\phi)$. Thus we must discard the possibility of $\mu\in\mathbb{Z}^+$ as well.

Upon carrying out the same analysis for $\text{LegendreQ}[\nu,\mu,x]$, it is once again found that its regularity requires $k=\text{imaginary}$ for both $\mu\in\mathbb{Z}^+$ as well as for $\mu=n+\frac{1}{2}=\text{half integer}, ~n\in \mathbb{Z}^+$. Moreover, if one considers $\text{LegendreP}[\nu,\mu,x]$ and $\text{LegendreP}[\nu,-\mu,x]$ as the complete set of basis with $\mu=n+\frac{1}{2}=\text{half integer}, ~n\in \mathbb{Z}^+$, then again the regularity condition forces $k=\text{imaginary}$.

Thus Ans\"atz I given by \ref{choice 1} is not an acceptable ans\"atz, since $\Theta(\theta)$  diverges at $\theta=\pi$ for both $\mu=\text{integer, non-integer}$. 

Therefore, we discard Ans\"atz I and proceed to analyse Ans\"atz II in the next subsection.

\subsubsection{Regularity Conditions for Ansatz II }
Taking Ans\"atz II given by equation \ref{choice 2} and choosing $c_6=0$, we can write the solution  $\Theta(\theta)$ expressed in equation \ref{theta2} as 
\begin{equation}
    \label{theta2-regularity}
   \begin{split}
       \Theta(x)&= c_5 (1-x^2)^{1/4}
   \text{LegendreP}[\frac{1}{2}, \frac{1}{2} \sqrt{1 -4 N + 4 k^2}, 
   x]
   \end{split}
\end{equation}
\begin{figure}[h!]
    \centering
    \includegraphics[width=0.8\linewidth]{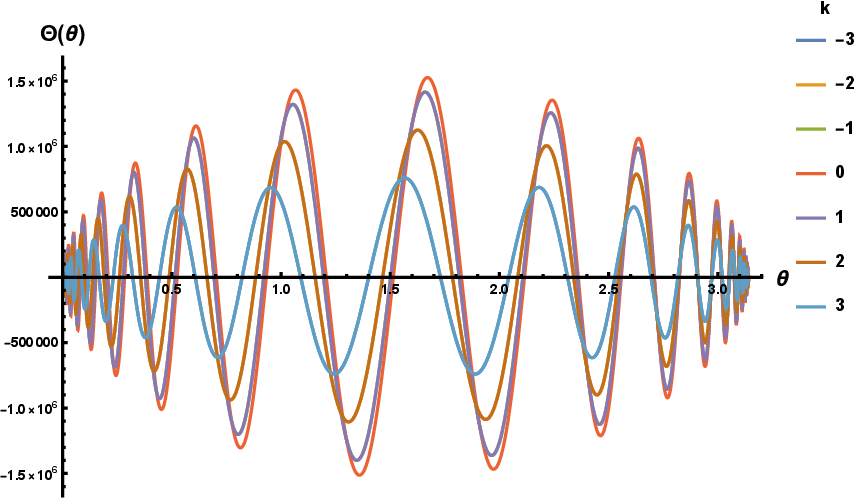}
    \caption{ Functional behaviour of the angular sector of the Goldstone mode $\Theta(\theta)$ for different values of $k$ with $N=10$.}
    \label{fig Theta vs theta}
\end{figure}
We note from equations \ref{P non-int mu, x=+1} and \ref{P non-int mu, x=-1} that for non-integer values of $\mu$, the asymptotic forms of the Legendre function are $P_\nu^\mu(x)\big|_{x\to\pm1}\sim(1\mp x)^{-\mu/2}$. Since $\mu=\frac{1}{2} \sqrt{1 -4 N + 4 k^2}$, the asymptotic forms of \ref{theta2-regularity} near the poles are 
\begin{equation}
    \label{final theta divergent x=+1}
    \Theta(x)\sim \big(1+x\big)^{\frac{1}{4}}\big(1-x\big)^{\frac{1}{4}[1-\sqrt{1 -4 N + 4 k^2}]}~~~x\to+1^-
\end{equation}
and
\begin{equation}
    \label{final theta divergent x=-1}
    \Theta(x)\sim \big(1-x\big)^{\frac{1}{4}}\big(1+x\big)^{\frac{1}{4}[1-\sqrt{1 -4 N + 4 k^2}]}~~~x\to-1^+~.
\end{equation}
For the regularity at both poles, equations \ref{final theta divergent x=+1} and \ref{final theta divergent x=-1} suggest $1-\sqrt{1 -4 N + 4 k^2}>0$. Thus, for non-integer $\mu=\frac{1}{2}\sqrt{1 -4 N + 4 k^2}$, we have the regularity condition 
\begin{equation}
    \label{condition on N1}
    N\geq k^2~.
\end{equation}

Moreover, if $\mu\in\mathbb{Z}^+$, then it is evident from equations \ref{P int mu, x=1} and \ref{P int mu, x=-1} that the Legendre functions in the limit $x\to\pm1$ behave like $P_\nu^\mu(x)\sim(1\mp x)^{\pm\mu/2}$. Therefore, in the limit $x\to\pm1$ the asymptotic forms for $\Theta(x)$ from equation \ref{theta2-regularity} are obtained as
\begin{equation}
    \label{final theta divergent x=+1 mu int}
    \Theta(x)\sim \big(1+x\big)^{\frac{1}{4}}\big(1-x\big)^{\frac{1}{4}[1+\sqrt{1 -4 N + 4 k^2}]}~~~x\to+1^-
\end{equation}
and
\begin{equation}
    \label{final theta divergent x=-1 mu int}
    \Theta(x)\sim \big(1-x\big)^{\frac{1}{4}}\big(1+x\big)^{\frac{1}{4}[1-\sqrt{1 -4 N + 4 k^2}]}~~~x\to-1^+~.
\end{equation}
Thus $\Theta(x)$ is regular as $x\to1^-$ whereas it is divergent in the limit $x\to-1^+$. 
Therefore, the regularity condition on $\Theta(x)$ is dictated by the second limit, yielding the {\em selection rule},
\begin{equation}
    \label{constraint on k}
    N\geq k^2~.
\end{equation}

Upon performing the same analysis on $\text{LegendreQ}[\nu,\mu,x]$, in is found that the above condition \ref{constraint on k} is sufficient for its regularity at both poles.

It remains to analyse the particular case of $N=k^2+\frac{1}{4}$, in which case $\mu=0$. Even though $\lim_{x\to1^-}P_{1/2}^0(x)=1$, the other asymptotic limit is $\lim_{x\to-1^+}P_{1/2}^0(x)\sim \ln(1+x)\to-\infty$.

Thus, $\lim_{x\to1^-}\Theta(x)\sim(1+x)^{1/4}\to\text{finite}$. On the other hand, for $x\to-1^+$, it is obvious that $\lim_{x\to-1^+}\Theta(x)\sim \lim_{x\to-1^+} (1-x^2)^{1/4}\ln(1+x)\to0$ because of the prefactor $(1+x)^{1/4}$ coming from $(1-x^2)^{1/4}$. Thus, the solution $\Theta(x)$ is regular at both poles.

The same analysis can be performed to show that regularity is maintained by $\text{LegendreQ}[\nu,0,x]$ at both poles. 

Therefore, the complete regular solution for $\Theta(\theta)$ is 
\begin{equation}
    \label{final theta sol}
   \begin{split}
       \Theta(\theta)&= c_5 (\sin\theta)^{1/2}
   \text{LegendreP}[\frac{1}{2}, \frac{1}{2} \sqrt{1 -4 N + 4 k^2}, 
   \cos\theta]\\
   &+ 
 c_6 (\sin\theta)^{1/2}
   \text{LegendreQ}[\frac{1}{2},\frac{1}{2} \sqrt{1 -4 N + 4 k^2}, 
   \cos\theta]
   \end{split}
\end{equation}
with the two parameters satisfying the conditions $N\in\mathbb{R}^+$, $k\in\mathbb{Z}$ and $N\geq k^2$ as obtained earlier in equation \ref{constraint on k}. 

Fig.~\ref{fig Theta vs theta} displays the functional behaviour of the angular sector of the Goldstone mode $\Theta(\theta)$ for different values of $k$ with $N=10$. It is clear from the figure that $\Theta(\theta)$ vanishes at the poles $\theta=0,\pi$, as also required for the regularity of the function $\Theta(\theta)$ at the poles. 

The selection rule $N\geq k^{2}$ has a direct physical interpretation beyond being a mathematical regularity requirement. The integer $k$ labels the azimuthal Fourier modes of the horizon Goldstone field, whereas $N$ is determined by the temporal dynamics through the ratio $N=f_1(v)/f_2(v)$ (Ans\"atz II) which is correlated with the black hole mass $m(v)$. Consequently, regularity at the poles imposes a mass-dependent upper bound on the azimuthal structure that can be supported by the Goldstone sector, that is, $|k|\leq \sqrt{N}$. Thus, the spectrum of admissible horizon supertranslation modes is not arbitrary: the geometry itself selects which azimuthal angular modes are physically allowed. This provides a non-trivial link between the macroscopic black hole parameter (mass) and the spectrum of its horizon soft degrees of freedom.

\begin{figure}[h!]
    \centering
    \includegraphics[width=0.8\linewidth]{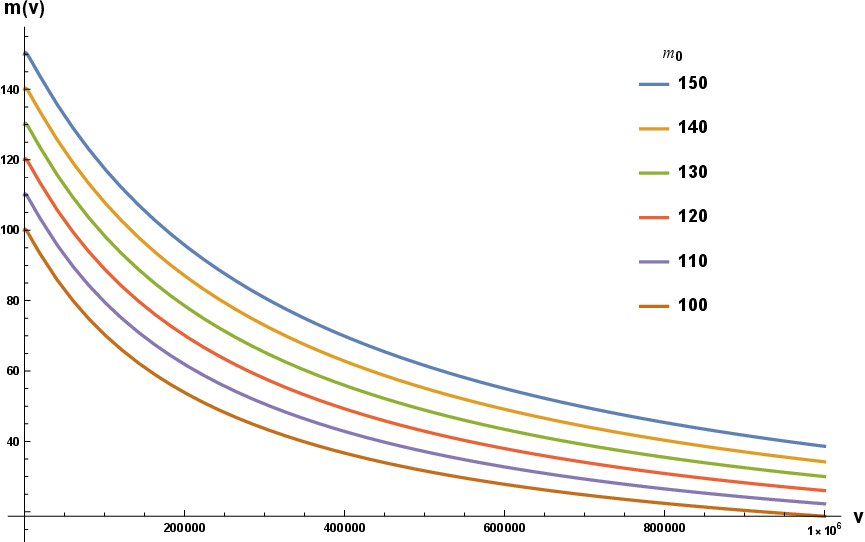}
    \caption{ Functional form of the mass function $m(v)$ for different values of initial black hole mass $m_0$ with the initial condition $m'(v=0)=-0.001$ in all cases.}
    \label{m vs v}
\end{figure}

\subsection{Analytical solution for the black hole mass}
In the preceding analysis, we have seen that Ans\"atz I gives diverging solutions whereas Ans\"atz II yields regular solutions at both poles with the condition \ref{constraint on k}, that is, $N\geq k^2$. Consequently, we adopt Ans\"atz II that implies $f_3=0$ given by equation \ref{choice 2}. From the last equation in \ref{KG coefficients}, we therefore have 
\begin{equation}
    \label{f(v)2}
    f(v)=\epsilon\Big[\frac{Cm^3}{m'} \Big]^{1/2}~.
\end{equation}
where $C$ is the integration constant and $\epsilon=\pm1$. 

Substituting $f(v)$ from equation \ref{f(v)2} in $f_1/f_2=N$ from Ans\"atz II (equation \ref{choice 2}), and using the expressions for $f_1$ and $f_2$ from \ref{KG coefficients}, we obtain
\begin{equation}
    \label{mass eqn}
    \begin{split}
      -12 N m^2 (m'')^2-4 (7 N-36) (m')^4&+4 (17 N-8) (m')^3+8 N m^2 \partial_v^3m m'\\
      &+(m')^2 \left\{16 (1-2 N) m m''+N\right\}=0
    \end{split}
\end{equation}
Since $m(v)$ changes very slowly due to Hawking radiation, $m''(v)$ and $m'''(v)$ can be neglected in the first approximation. This leads to
\begin{equation}
    \label{1st approx}
    -4 (7 N-36) (m')^4+4 (17 N-8) (m')^3+N(m')^2=0
\end{equation}
This is a fourth order {\em algebraic} equation in $m'$ that gives four constants as its roots. Two of them are zero and the rest of the two roots are non-zero constants. Considering any of the constant root $m'=A$, its solution is $m=Av+B$, where  $B$ is the integration constant.  This approximate solution requires improvement when we do not neglect $m''$ and $m'''$. Such improved solution  can be expressed  through a Pad\'e approximant for the mass function as
\begin{equation}
    \label{mass sol}
    m(v)=\frac{a_0+a_1v}{1+b_1v+b_2v^2}~.
\end{equation}

Substitution of equation \ref{mass sol} in equation \ref{mass eqn}, and denoting the initial conditions with $m(v=0)=m_0=a_0$ and $m'(v=0)=-\beta$, the constants appearing in \ref{mass sol} are determined consistently, yielding
\begin{equation}
    \label{a,b}
    \begin{split}
        &a_1=-\frac{\beta}{\Sigma}  \Big[256 \beta ^2 \left(\Delta -4 \beta ^2\right)+\left(12800 \beta ^4+124440 \beta ^3-64284 \beta ^2+1530 \beta -9\right) N^4\\
        &-2 N^3 \big\{113360 \beta ^4-71904 \beta ^3-4 \beta ^2 (185 \Delta +7041)-60 \beta  (119 \Delta -6)+99 \Delta        \big\}\\
        &-12 N^2 \big\{14240 \beta ^4+6440 \beta ^3+4 \beta ^2 (574 \Delta +293)+798 \beta  \Delta -3 \Delta \big\}\\
        &-32 \beta  N \left(956 \beta ^3+264 \beta ^2-147 \beta  \Delta -42 \Delta \right)   \Big]=\mathcal{O}(m_0^0)~~,\\
        &b_1=-\frac{\beta}{m_0\Sigma}  \Bigg[256 \beta ^2 \left(\Delta -4 \beta ^2\right)+\left(5600 \beta ^4+51000 \beta ^3-63204 \beta ^2+1530 \beta -9\right) N^4\\
        &-16 N^3 \big\{5800 \beta ^4-12066 \beta ^3-7 \beta ^2 (10 \Delta +501)+\beta  (45-663 \Delta )+9 \Delta     \big\}\\
        &-12 N^2 \big\{15968 \beta ^4+7016 \beta ^3+4 \beta ^2 (436 \Delta +293)+654 \beta  \Delta -3 \Delta \big\}\\
        &-64 \beta  N \left(496 \beta ^3+132 \beta ^2-78 \beta  \Delta -21 \Delta \right)\Bigg]=\mathcal{O}(m_0^{-1})~~,\\
        b_2&=-\frac{\beta ^2}{2 m_0^2\Sigma} \Bigg[128 \beta ^2 \left(\Delta -4 \beta ^2\right)+\left(2800 \beta ^4+28560 \beta ^3-480 \beta ^2-612 \beta +9\right) N^4\\
        &+4 N^3 \big\{-11600 \beta ^4+9696 \beta ^3+28 \beta ^2 (5 \Delta +3)+12 \beta  (85 \Delta +6)-9 \Delta    \big\}\\
        &-48 \beta  N^2 \left(1996 \beta ^3+376 \beta ^2+218 \beta  \Delta +\beta +57 \Delta \right)-64 \beta  N \big(248 \beta ^3+48 \beta ^2-39 \beta  \Delta -6 \Delta \big)\Bigg]=\mathcal{O}(m_0^{-2}),
    \end{split}
\end{equation}
where
\begin{equation}
    \label{Delta}
    \Delta=\sqrt{\beta ^2 \left(16 \beta ^2-20 \beta ^2 N^2-204 \beta  N^2+3 N^2+368 \beta ^2 N+96 \beta  N\right)}
\end{equation}
and 
\begin{equation}
    \label{Sigma}
    \Sigma=18 N \big\{-16 \beta ^2+\left(20 \beta ^2+204 \beta -3\right) N^2-16 \beta  (23 \beta +6) N     \big\} \big\{\Delta +4 \beta ^2 (5 N-1)\big\}.
\end{equation}

Figure \ref{m vs v} shows the dependency of the mass function $m(v)$ on the retarded time $v$ for different values of initial mass $m_0$. Due to Hawking evaporation, we consider $m'(v=0)=-\beta$  with $\beta=0.001$.  The mass function shows a monotonic decreasing nature with $v$ and the sign of $m'(v)$ remains negative implying persistence of Hawking evaporation throughout the evolution. Moreover, it is evident from the figure  that higher mass black holes take longer times compared to the lower ones to evaporate completely, which is consistent with the expected property of Hawking evaporation~\cite{ hawking1974black, Hawking1975}.

\begin{figure}[h!]
    \centering
    \includegraphics[width=0.7\linewidth]{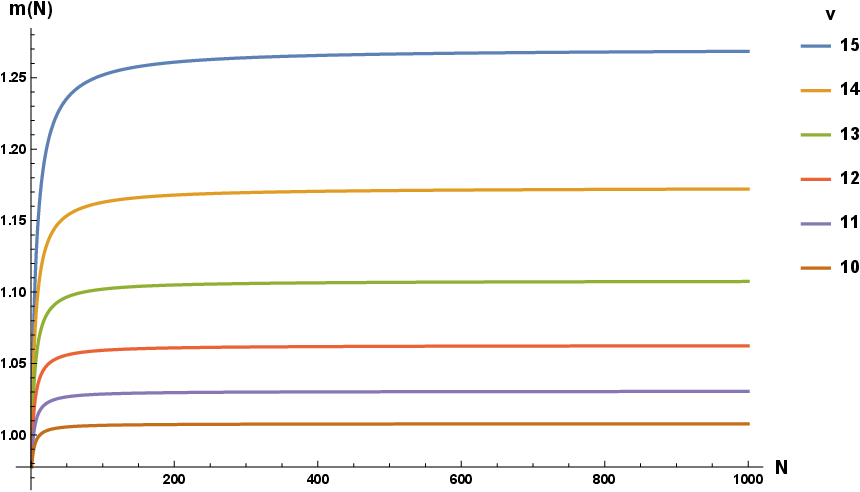}
    \caption{Illustration of the behaviour of $m(N)$ on different constant-time hypersurfaces $v = \text{constant}$. The black hole mass function increases monotonically with $N$.}
    \label{m VS N plot}
\end{figure}

For the set of constants specified above, Fig. \ref{m VS N plot} demonstrates that, on constant-time hypersurfaces $v=\mathrm{constant}\neq0$, the black hole mass function increases monotonically with the proportionality constant $N$. Although $k^2\in\mathbb{Z}^+$, the parameter $N$ is not restricted to discrete values and can vary continuously over $\mathbb{R}^+$. Consequently, the mass function exhibits a continuous dependence on $N$. The corresponding increase in $N$ is also accompanied by an enlargement of the event horizon. Since the allowed azimuthal modes of the Goldstone field are constrained by $N\geq k^2$, a larger value of $N$ permits progressively higher-order $k$-modes consistent with the regularity condition in the $\theta$-sector. Thus, a bigger black hole with a larger horizon is accompanied with a larger number of azimuthal modes, $k^2_\text{max}=N$, that can be consistently supported on the horizon.

An important dynamical implication follows from the constraint $k^2_\text{max}=N$ in conjunction with Fig. \ref{m VS N plot}. A black hole with a relatively large initial mass, or equivalently a large value of $N$, can support a larger set of azimuthal $k$-modes of the Goldstone field on its horizon. As the black hole undergoes Hawking evaporation, its mass decreases and the event horizon correspondingly shrinks. This evolution reduces the allowed range of $k$ according to the condition $N\geq k^2$, thereby progressively excluding higher-order azimuthal modes while retaining only the lower-order modes compatible with the instantaneous size of the horizon. Consequently, modes that cease to satisfy the regularity condition are no longer supported by the evolving horizon and are naturally associated with degrees of freedom transferred away from the horizon during the evaporation process. The evaporation therefore induces a dynamical reorganization of the horizon soft sector, with the set of admissible Goldstone modes evolving continuously as the black hole mass decreases. This provides a physical interpretation of the evaporation process as a dynamical filtering mechanism for the horizon modes, whereby higher-order azimuthal modes are successively removed as the horizon contracts.

\subsection{Solution for temporal part of Goldstone field}
Substituting the mass function $m(v)$ from equation \ref{mass sol} in the expression for the temporal part $f(v)$ given by \ref{f(v)2}, we obtain the temporal part of the Goldstone field as
\begin{equation}
    \label{goldstone v sol}
    \begin{split}
        f(v)=\epsilon  \sqrt{-\frac{C \left(a_1 v+m_0\right){}^3}{\left(b_2 v^2+b_1 v+1\right) \left[a_1 \left(b_2 v^2-1\right)+m_0 \left(2 b_2 v+b_1\right)\right]}}
    \end{split}
\end{equation}

Thus, the complete solution for the Goldstone modes, namely $F(v,\theta,\phi)=f(v)\Theta(\theta)\Phi(\phi)$, from equations \ref{Phi}, \ref{final theta sol}, \ref{goldstone v sol}, residing on the horizon $r=0$ of a Vaidya-Schwarzschild black hole, turns out to be
\begin{equation} 
    \label{complete f sol}
    \begin{split}
        F&(v,\theta,\phi)=\epsilon  \sqrt{-\frac{C \left(a_1 v+m_0\right){}^3}{\left(b_2 v^2+b_1 v+1\right) \left[a_1 \left(b_2 v^2-1\right)+m_0 \left(2 b_2 v+b_1\right)\right]}}\\
        &\times\Bigg[c_5 (\sin\theta)^{1/2}
   \text{LegendreP}[\frac{1}{2}, \frac{1}{2} \sqrt{1 -4 N + 4 k^2}, 
   \cos\theta]\\
   &+ 
 c_6 (\sin\theta)^{1/2}
   \text{LegendreQ}[\frac{1}{2},\frac{1}{2} \sqrt{1 -4 N + 4 k^2}, 
   \cos\theta] \Bigg]\big[c_1e^{ik\phi}+c_2e^{-ik\phi}\big]
    \end{split}
\end{equation}
Equation \eqref{complete f sol} demonstrates that the horizon Goldstone field inherits its temporal behaviour from the evolving black hole mass $m(v)$ represented by the Pad\'e approximant. The angular dependence is constrained by the regularity condition selected in the Legendre sector. The resulting solution therefore describes a dynamic configuration rather than an independently prescribed Goldstone field on a fixed background. Importantly, evolution of the black hole mass directly changes the temporal behaviour of the horizon Goldstone excitation, providing an explicit realization of the backreaction on the horizon soft degrees of freedom.

\section{Discussion and Conclusion}\label{disc}
The emergence of Goldstone modes as a consequence of spontaneous symmetry breaking is a well-established phenomenon in quantum field theory and critical phenomena~\cite{Zinn-Justin}. In the context of black hole spacetimes, an analogous structure arises when horizon-preserving {\em large} diffeomorphisms act non-trivially on the gravitational configuration. In the present work, we investigated this scenario for a dynamical Vaidya–Schwarzschild black hole and showed that the associated horizon supertranslation degree of freedom can be treated as a dynamical Goldstone mode. Unlike a fixed background analysis, the time dependence of the Vaidya mass introduces a non-trivial coupling between the horizon soft sector and the evolving black hole horizon. This provides a setting in which the dynamics of horizon degrees of freedom can be studied simultaneously with black hole evaporation.

A notable feature of the resulting equations is that the horizon Goldstone sector and the black hole evolution are dynamically connected intimately. However, these equations form a high nonlinear system of the coupled differential equations and obtaining an exact closed-form solution is not feasible. We therefore looked for an approximate analytical solution that captures the {\em essential physics}  of the coupled dynamics. Accordingly, we first considered a separable form for the Goldstone field and made two possible ans\"atze. One of the ans\"atze turned out to be inadmissible while the other was found to give regular solution in the angular sector of the Goldstone field.

Noting that the black hole mass decreases slowly, we employed the slowly varying approximation followed by a Pad\'e approximant of order $(1,2)$ for the mass function. The resulting solutions reproduce the expected qualitative behaviour of an evaporating black hole: for appropriate initial conditions, the mass decreases monotonically with the evolution. The solutions also show that more massive black holes evaporate over longer timescales than less massive ones, consistent with the characteristic behaviour expected for Hawking evaporation.

The corresponding temporal Goldstone solution is determined by the evolving mass function and is therefore not independent of the black hole dynamics. As the mass changes, the temporal prefactor of the Goldstone field evolves accordingly, while its angular dependence remains constrained by the regularity-selected Legendre sector. The complete solution thus captures the essential physics and describes a horizon excitation whose temporal behaviour is coupled to the evolution of the macroscopic parameter $m(v)$. This provides an explicit realization of the idea that horizon soft degrees of freedom should evolve together with the black hole rather than remain as static labels attached to a fixed horizon.

It is important to note that the periodicity of the azimuthal sector requires the Fourier mode $k$ to be an integer, while regularity at the two poles places an additional restriction on the allowed Legendre sector. In particular, the first possible ans\"atz, $f_{1}=0$, is incompatible with regularity of the angular modes at the poles. The second ans\"atz, $f_{3}=0$, on the other hand, admits regular solutions providing a condition on the maximally allowed azimuthal mode,  $k^2_\text{max}=N$.  Since $k$ labels the azimuthal structure of the horizon Goldstone modes, while $N$ is related to the black hole mass through the temporal dynamics, the regularity of the $\theta$-modes establishes a direct relation between the macroscopic property of the black hole and the spectrum of its soft gravitational excitations. Thus, the black hole mass itself determines which members of the angular sector of the  Goldstone modes can be supported.

A central aspect of our analysis is that the Goldstone dynamics is obtained directly from the Einstein–Hilbert action. The metric generated by the horizon-BMS transformation is expanded around the unperturbed Vaidya background, and the quadratic fluctuation sector provides the effective action governing the Goldstone field. Consequently, the Goldstone mode is not introduced as an independent matter field but arises from the gravitational degrees of freedom associated with the non-trivial action of the {\em large} diffeomorphism. The resulting equations of motion demonstrate that the two sectors are intrinsically coupled: the evolving mass function $m(v)$ enters the Goldstone equation, while the Goldstone configuration $F(v,\theta,\phi)$ contributes to the equation governing the evolution of $m(v)$. The horizon soft sector therefore participates in the dynamical description of the black hole rather than simply labelling a family of otherwise equivalent configurations.

The constraint on the azimuthal modes acquires particular significance when the black hole is dynamical. During evaporation, the decreasing black hole mass is accompanied with decrease in $N$, and the regularity condition in the $\theta$-sector of the Goldstone field forces the allowed values of $k$ in progressively smaller ranges, restricted by $k^2\leq N$. A massive black hole supporting a rich spectrum of horizon Goldstone excitations, in the process of ongoing evaporation, is restricted to progressively fewer low-$|k|$ modes. Consequently, the evaporation process induces a dynamical filtering of the horizon soft sector. The important point is that this filtering originates from the regularity of the Goldstone modes living on the evolving horizon rather than from imposing an external cutoff on the Goldstone spectrum. The number of horizon-supported modes is consequently not fixed throughout the evolution but is correlated with the instantaneous mass of the black hole.

Overall, the present work provides an effective field theory framework in which near-horizon BMS symmetry breaking, Goldstone dynamics, and black hole evaporation can be treated within a single gravitational description. The most significant consequence is the emergence of a mass-dependent selection rule for the horizon soft spectrum and its consequent evolution during evaporation. This establishes a concrete connection between the macroscopic dynamics of an evolving black hole and dynamical organisation of the spectrum of its horizon soft degrees of freedom. Further investigation of the entropy carried by the allowed modes, their late-time dynamics, and their possible relation to information transfer can provide a promising future route toward understanding the microscopic role of horizon symmetries in black hole thermodynamics.

\section*{Acknowledgement} 
Nihar Ranjan Ghosh is supported through a Research Fellowship from the Ministry of Human Resource Development (MHRD), Government of India.


\end{document}